\documentclass{article}

\usepackage{amsmath}
\usepackage{amssymb}
\usepackage{amsthm}
\usepackage{graphicx}
\usepackage{comment}
\usepackage{authblk}
\usepackage{afterpage}

\title{Why we like the ECI+ algorithm}
\author[1]{Andrea Gabrielli}
\author{Matthieu Cristelli}
\author[2]{Dario Mazzilli}
\author[1]{Andrea Tacchella}
\author[1]{Andrea Zaccaria}
\author[1,2]{Luciano Pietronero\thanks{luciano.pietronero@roma1.infn.it}}

\affil[1]{Institute for Complex Systems, CNR}
\affil[2]{Department of Physics, Sapienza University of Rome}

\newcommand{\be}{\begin{equation}}
\newcommand{\ee}{\end{equation}}
\newcommand{\bea}{\begin{eqnarray}}
\newcommand{\eea}{\end{eqnarray}}

\begin{document}

\newtheorem{lem}{Lemma}
\newtheorem{thm}{Theorem}
\newtheorem{cor}{Corollary}
\thispagestyle{empty}
\maketitle
\abstract{Recently a measure for Economic Complexity named ECI+ has been proposed by Albeaik et al. We like the ECI+ algorithm because it is mathematically identical to the Fitness algorithm, the measure for Economic Complexity we introduced in 2012. We demonstrate that the mathematical structure of ECI+ is strictly equivalent to that of Fitness (up to normalization and rescaling). We then show how the claims of Albeaik et al. about the ability of Fitness to describe the Economic Complexity of a country are incorrect. Finally we hypothesize how the wrong results reported by these authors could have been obtained by not iterating the algorithm.}

\section{From the Fitness algorithm to ECI+ with a simple relabeling}
Let us call $X_{cp}$ the {\em extensive} export matrix giving, in a fixed year, the export expressed in dollars of the product $p$ by the country $c$.
By its definition we have that
\[X_c=\sum_p X_{cp}\]
is the total export of country $c$ in that year.
Analogously the quantity
\[X_p=\sum_c X_{cp}\]
gives the total amount (in dollars) of product $p$ exported in the same year by all countries.
Finally, we call 
\[X=\sum_{cp}X_{cp}\]
the total world export of the considered year.\\
Let us now recall the fundamental equations of the algorithm \cite{scirep2012, plos2013} from which it is possible to compute the Fitness of countries and the Complexity of products from the COMTRADE data of international export.\\
In this case the export matrix is binarized by using the Revealed Comparative Advantage (RCA) criterion. The RCA of a country $c$ on a product $p$ is defined as
\[RCA_{cp}=\frac{X_{cp}/X_c}{X_p/X}\]
which can be read as the ratio between the share of product $p$ in the export basket of country $c$ and the share of the same product in the total world export (or equivalently as the ratio between the share of the country $c$ in the total export of product $p$ and the share of the same country in the total world export). 
$RCA_{cp}$ is generally considered to give a measure of how ``good" is a country $c$ in exporting (and therefore producing) a product $p$: if $RCA_{cp}>1$ the country $c$ is in average better than the rest of the world to export $p$.
Consequently, the criterion to introduce the {\em binary} export matrix $M_{cp}$ is simply the following: if $RCA_{cp}>1$ then $M_{cp}=1$, while if $RCA_{cp}\le 1$ then $M_{cp}=0$. \\
Through the matrix $M_{cp}$ we can define the fitness-complexity algorithm respectively for the fitness $F_c$ of countries and complexity $Q_p$ of products as  \cite{scirep2012,plos2013}
\be
\left\{
\begin{array}{l}
F_c^{(N+1)}=\sum_p M_{cp} Q_{p}^{(N)}\\
\\
Q_{p}^{(N+1)}=\frac{1}{\sum_c\frac{M_{cp}}{F_c^{(N)}}}
\end{array}
\right.
\label{fit-com2}
\ee
with the condition of normalizing at each step all $F_c$'s and $Q_p$'s by dividing at each iteration their values by the mean values respectively over all $c$ and all $p$ at the same iteration in order to avoid possible divergences due to the hyperbolic nature of the second equation.\\
We now show in few steps that ECI+ and PCI+ formulas defined in \cite{ECI+} can be simply seen as the version of Eqs.~\eqref{fit-com2} where $M_{cp}$ is simply substituted by the extensive matrix $X_{cp}$ (change that was already discussed in \cite{plos2013}).\\
First, let us substitute the second equation of \eqref{fit-com2} at iteration $N$ in the first one at iteration $N+1$:
\be
F_c^{(N+1)}=\sum_p M_{cp}\frac{1}{\sum_{c'}\frac{M_{c'p}}{F_c'^{(N-1)}}}
\label{fit-com3}
\ee
If we now substitute $X_{cp}$ to $M_{cp}$ and rename $F_c^{(2N)}=X_c^{N}$ we get exactly Eq.~(10) in \cite{ECI+}:
\be
X_c^{N}=\sum_p X_{cp}\frac{1}{\sum_{c'}\frac{X_{c'p}}{X_c'^{N-1}}}
\label{eci+1}
\ee
In order to rank countries the authors of \cite{ECI+} propose a measure of competitiveness of countries, called ECI+, which, a part from the subtraction of an iteration independent offset\footnote{This country-dependent offset $\log \frac{\sum_c X_{cp}}{X_p}$ can be seen as obtained by the same formula \eqref{eci+1} for $X_c^N$ with $X_c^{N-1}=1$ for all $c$. Note that, differently from what written in \cite{ECI+}, the argument of the logarithm is not the ``average share that the country represents in the
export of a product", but the {\em sum} of the shares of all countries in the export of the product $p$.},  is given by  $\log X_c^\infty$ in strict analogy with what was done for instance in \cite{plos2015,plos2017}.
Analogously, if (i) we substitute the first equation of Eq.~\eqref{fit-com2} at iteration $N$ in the second one, (ii) substitute $X_{cp}$ to $M_{cp}$ in the same equation, and rename $1/Q_{p}^{(2N)}=X_p^N$ we get exactly Eq.~(13) of \cite{ECI+}:
\be
X_p^{N}=\sum_c X_{cp}\frac{1}{\sum_{p'}\frac{X_{cp'}}{X_p'^{N-1}}}
\label{pci+1}
\ee
The reciprocal algebraic relation between $Q_p^{(2N)}$ and $X_p^N$ is recovered in the definition of the metrics called PCI+ in Eq.~16 in \cite{ECI+} as $-\log X_p^\infty=\log(1/ X_p^\infty)$, apart from the addition of another iteration independent offset\footnote{We do not understand why the authors of \cite{ECI+} in the definition $[\log X_p-\log X_p^\infty]$ of PCI+ have $X_p^\infty$ which is adimensional while $X_p$ is measured in dollars. Why does this metrics change if we measure export in euros instead of dollars?} $\log X_p$. \\
Similarly to the fitness-complexity algorithm, both $X_c^N$ and $X_p^N$ are normalized at each iteration by dividing by an appropriate mean of their values respectively over all countries and all products in order to avoid divergences due to the non-linear hyperbolic nature of the algorithm. The authors chose for this purpose the geometric mean, probably taking into account the extensive nature of the matrix $X_{cp}$.

\section{Problems and inconsistencies in fitness results}
Given the equivalence of the algorithms, the claim reported in \cite{ECI+} that continuous data can be used in ECI+ but not in Fitness is indeed extravagant. In the second part of \cite{ECI+} it is argued that the same algorithm works well when it is named ECI+ but not when it is named Fitness.\\
The solution of the puzzle is in the different input data used. Clearly, the Fitness algorithm can be used with continuous, discrete, intensive, or extensive data, depending on the objective of the analysis, as already discussed in \cite{plos2013}. 
Albeaik et al. mix different input data (extensive for ECI+ and intensive for Fitness), and this is used as an erroneous evidence for an apparent difference in the algorithms. Moreover they state that Fitness is strongly correlated with diversity, as it is obvious due to the explicit sum over the products that a country exports. This sum however is weighted by the complexity of products, and this introduces residuals that are strongly informative. However the diversity term is important and cannot be disregarded, as it is a fundamental principle of Economic Complexity. Being defined in the same exact way, ECI+ also has a dependence on country size, which is trivially and explicitly removed by subtracting the term $\log(\sum_p\frac{X_{cp}}{X_p})$, which has a 0.97 correlation with $X_c^\infty$.\\
The Fitness measure as reported in \cite{ECI+} shows an anomalous ranking, in sharp contrast with the established literature \cite{plos2013}. In order to investigate this puzzle we reconstructed the input data used in that paper to the best of our ability and ran both algorithms. The results are discussed in the following.\\
In \cite{ECI+} it is stated that "Fitness Complexity ranks many Southern European countries (such as Spain, Italy, and Portugal) at the top of the ranking, and also, provides very low complexity values for advanced East Asian and European economics, such as South Korea, Switzerland, Finland, Japan, and Singapore".
First, it should be noted that country rankings obtained with Fitness have been published in \cite{plos2013}, and ignored by Albeaik et al., who report very different results, obtained only by them to the best of our knowledge. Indeed, the real Fitness shows that the top 5 countries by Fitness in 2010 were Germany, China, Italy, Japan and United States; moreover, the 5 countries that are said to have "very low complexity values" are all in the top 20\% of the ranking \cite{plos2013}. One should note that these datasets refer only to manufacturing, without taking into account services. \\
Moreover, Albeaik et al. state that "for the Fitness measure, the economy of Greece is ranked higher than that of Japan, Sweden, or China." This is again wrong and inconsistent with the results we published in  \cite{plos2013}, where Greece is ranked 34th, while Japan, Sweden, and China are 4th, 14th, and 2nd respectively. We point out that, given the strong weight the Fitness measure gives to diversification, it is really hard to believe a dataset exists such as China is ranked below Greece.\\
The strong differences between the rankings published in \cite{plos2013} and those reported in \cite{ECI+} can not be explained by a difference in the starting dataset alone. We base our following analysis on the BACI dataset \cite{BACI}, which we filter following the prescriptions given in \cite{ECI+}. On such dataset we recompute Fitness and ECI+ up to convergence, following the prescriptions given in \cite{conv} (we give the number of the performed iterations for reproducibility: 200). Curiously enough, the Fitness rankings reported in \cite{ECI+} are extremely similar to those that one would obtain on the same dataset by iterating the algorithm \textit{just once}, which appears totally unreasonable.  If one iterates the Fitness algorithm for 200 steps, the rankings appear much more reasonable and very different from those reported in the ECI+ paper, as well as coherent with those reported in \cite{plos2013}.\\
We found also very strange that, in order to have Spain at the top of the Fitness ranking, as reported by Albeaik et al, after 1 iteration we have to set $F_c^0 = k_c^0$ and $Q_p^0=k_p^0$ instead of the usual constant initial conditions used for Fitness ($F_c=1 \forall c$ and $Q_p=1 \forall p$). While the starting point of the iteration procedure becomes irrelevant when the algorithm is iterated up to convergence (and in this case Spain is never at the top), obviously it becomes more important if the number of iterations is reduced. In order to understand which misconceptions lead Albeaik et al. to such Fitness ranking, we tried to reproduce their results by visually comparing our computations of the Fitness and ECI+ with theirs. In particular, in fig. 1 we reproduced the original comparison between ECI+ and Fitness as presented in \cite{ECI+} and compare it with the one recomputed by us\footnote{In order to reproduce the figures in \cite{ECI+} we standardized the Fitness. We point out that the correct 
procedure is to take its logarithm.}. The upper figure is the original Fitness vs ECI+ graph taken from \cite{ECI+}, the center figure is our best reproduction of that plot, that is done by performing only one (1) iteration of Fitness and with the initial conditions mentioned above. In the lower panel we present the comparison between ECI+ and the logarithm of the value of Fitness at convergence, obtained with 200 iterations. While the authors of \cite{ECI+} compute the logarithm of $X$ after iterating, they omit to compute the logarithm of Fitness as it is usually done in the literature \cite{scirep2012,plos2013,plos2015,plos2017} to compare it to other macroeconomic intensive indicators. One can easily realize that the best reproduction of the results presented by Albeaik et al. is obtained if the Fitness algorithm is iterated only once, which is clearly a mistake. On the contrary, if the same algorithm is iterated up to convergence, the two measures correlate more, given the mathematical equivalence of 
the algorithms. However, since the input matrices differ, some deviations are still present.
\section*{Conclusions}
In summary, the paper of Albeaik et al. \cite{ECI+} does not introduce any new algorithm but just renames the Fitness one as "ECI+''. In this respect one may also note that ECI+ has nothing to do with the original ECI \cite{ECI}. A detailed discussion of the problems of ECI and the reasons to introduce the Fitness can be found in \cite{plos2013}. Also in that occasion the authors of \cite{ECI} took inspiration from our work and learned, without ever citing, that the linear calculation of the ECI can be solved exactly by computing an eigenvector rather than with 18 iterations.\\
The numerical results reported in \cite{ECI+} for the Fitness are incorrect and even embarrassing in view of the mathematical equivalence of ECI+ and Fitness. Albeaik et al. \cite{ECI+} present a totally distorted view of the situation, from both a mathematical and numerical point of view.

\afterpage{ 
\thispagestyle{empty}

\begin{figure}
\label{cfr}
\centering
\includegraphics[scale=1.]{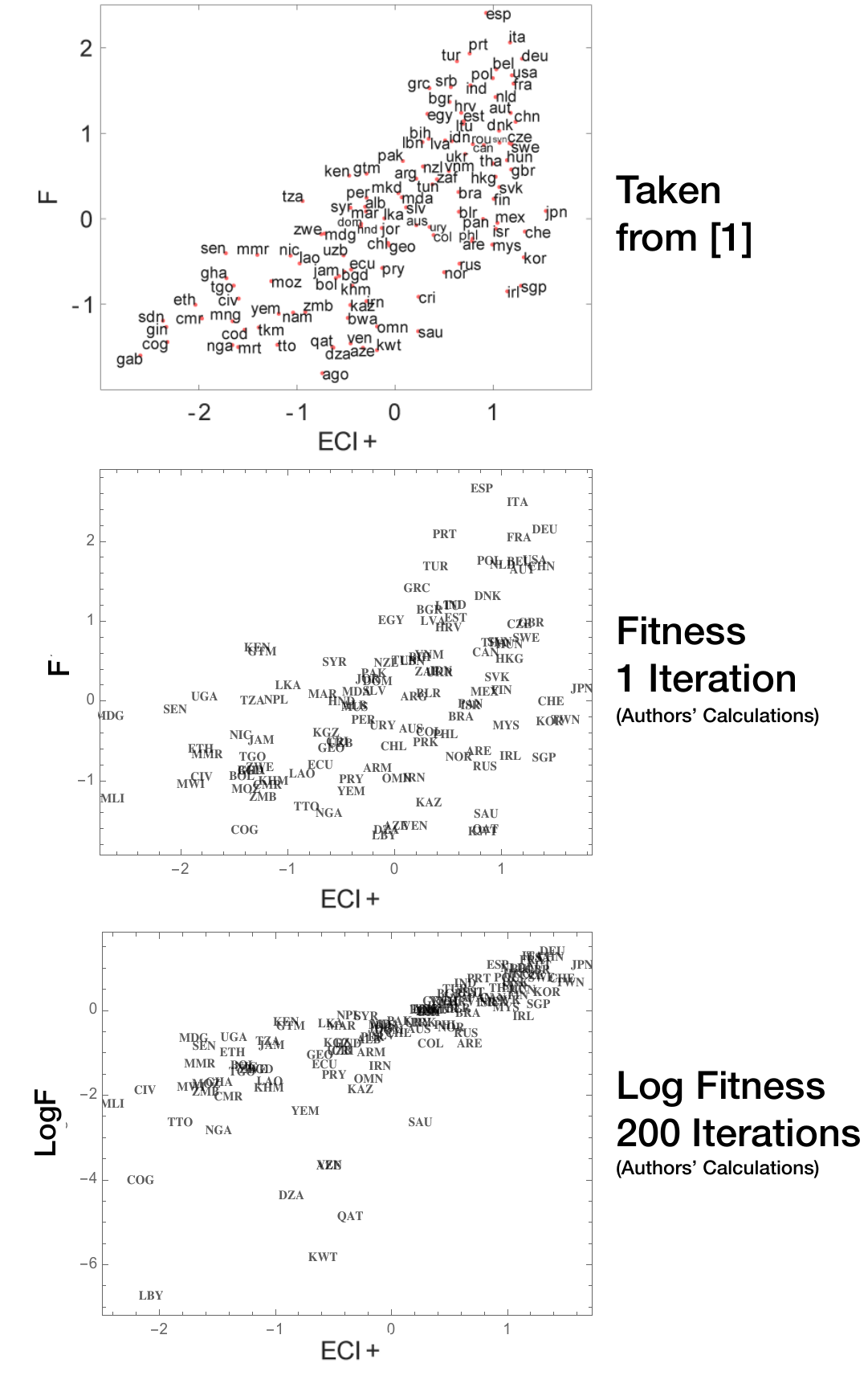}
\caption{Upper figure: the original comparison between Fitness and ECI+, as presented in \cite{ECI+}. Central figure: our best attempt to reproduce the above results: the Fitness algorithm is iterated only once. Lower figure: when the Fitness algorithm is iterated up to convergence, the two measures became highly correlated, as one may expect from the identical mathematical structure of the algorithms. The residual differences are due to the different input matrix and the normalizations.}
\end{figure}
}

\begin{thebibliography}{99}
\bibitem{scirep2012} A. Tacchella, M. Cristelli, G. Caldarelli, A. Gabrielli, L. Pietronero, Scientific Reports {\bf 2}, 723 (2012).
\bibitem{plos2013} M. Cristelli, A. Gabrielli, A. Tacchella, G. Caldarelli, L. Pietronero, PLoS ONE {\bf 8}, e7072 (2013).
\bibitem{ECI+} S. Albeaik, M. Kaltenberg, M. Alsaleh, C. A. Hidalgo, https://arxiv.org/abs/1707.05826v3 (2017).
\bibitem{plos2015} M. Cristelli, A. Tacchella, L. Pietronero, PLoS ONE {\bf 10}(2), e0117174 (2015).
\bibitem{plos2017} E. Pugliese, G.L. Chiarotti, A. Zaccaria, L. Pietronero, PloS ONE {\bf 12}(1), e0168540 (2017)
\bibitem{BACI} G. Gaulier, and S. Zignago. "Baci: international trade database at the product-level (the 1994-2007 version)." (2010).
\bibitem{conv} E. Pugliese, A. Zaccaria, and L. Pietronero. The European Physical Journal Special Topics 225.10: 1893-1911. (2016)
\bibitem{ECI} C.A. Hidalgo and R. Hausmann. PNAS 106.26: 10570-10575 (2009).
\end{thebibliography}
\end{document}